\documentclass[aps,prd]{revtex4}
\usepackage{subfigure,graphics} 
\usepackage{flafter} 
\begin{document} 

\title{
Non-gaussianity in fluctuations from warm inflation} 
\author{Ian G. Moss}
\email{ian.moss@ncl.ac.uk}
\author{Chun Xiong}
\affiliation{School of Mathematics and Statistics, University of  
Newcastle Upon Tyne, NE1 7RU, UK}

\date{\today}


\begin{abstract}
The scalar mode density perturbations in a the warm inflationary scenario
are analysed with a view to predicting the amount of non-gaussianity
produced by this scenario. The analysis assumes that the inflaton evolution is 
strongly damped by the radiation, with damping terms that are temperature 
independent. Entropy fluctuations during warm inflation play a crucial role in
generating non-gaussianity and result in a distinctive signal which should be
observable by the Planck satellite.

\end{abstract}
\pacs{PACS number(s): }

\maketitle
\section{introduction}

Observations of the cosmic microwave background are consistent with the
existence of gaussian, weakly scale dependent, density perturbations as
predicted by most inflationary models \cite{Spergel:2006hy}. 
The amount of non-gaussianity produced
by the simplest inflationary models is small and unlikely to be to be
observable by the next generation of experiments, but this still leaves open
the possibility that a slightly more exotic inflationary model could produce a
measurable effect.

One variation on inflation is the warm inflationary scenario \cite{berera95}
(see also \cite{moss85}). Warm inflation is characterised by the amount of
radiation production during the inflationary era. If the radiation field is in
a highly excited state during inflation, and this has a strong damping effect
on the inflaton dynamics, then we have what is know as the strong regime of
warm inflation. In this case, thermal fluctuations in the radiation are
transfered to the inflaton   
\cite{berera96,Lee:1999iv,berera97,berera98,berera00,taylor00} 
and become the primary source of
density fluctuations.

The possibility of warm inflation occurring in realistic particle models has
been enhanced by the discovery of a decay mechanism which is present in many
supersymmetric theories 
\cite{berera02,Berera:2004kc,Bastero-Gil:2005gy}. 
In these models,
the inflaton decays into light radiation fields through a heavy particle
intermediary. If the coupling constants are sufficiently large, these models
can lead to warm inflation.

Fluctuations in the cosmic microwave background allow us to measure the density
fluctuations at the surface of last scattering. We know, in principle, how to
evolve these fluctuations from early times using, for example, the Bardeen
variable $\zeta$ \cite{Bardeen:1980kt}. Observations can be compared to
predictions for various
moments of the probability distribution of $\zeta$. The most important of
these is the primordial power spectrum of fluctuations $P_\zeta(k)$, defined
by the stochastic average
\begin{equation}
\langle \zeta({\bf k}_1)\zeta({\bf k}_2)\rangle=(2\pi)^3P_\zeta(k_1)
\delta^3({\bf k}_1+{\bf k}_2).
\end{equation}
The bispectrum $B_\zeta(k_1,k_2,k_3)$, defined by 
\begin{equation}
\langle \zeta({\bf k}_1)\zeta({\bf k}_2)\zeta({\bf k}_3)\rangle=
(2\pi)^3B_\zeta(k_1,k_2,k_3)
\delta^3({\bf k}_1+{\bf k}_2+{\bf k}_3),
\end{equation}
can be used to examine the non-gaussianity in the density fluctuations 
The normalised amount of non-gaussianity in the bispectrum is
described by a non-linearity function $f_{NL}$, defined by
\begin{equation}
f_{NL}(k_1,k_2,k_3)={5\over 6}{B(k_1,k_2,k_3)\over
P(k_1)P(k_2)+P(k_2)P(k_3)+P(k_3)P(k_1)}.
\label{fdef}
\end{equation}
where the $5/6$ factor is convenient for cosmic microwave background
comparisons \cite{Komatsu:2001rj}.

Non-gaussianity can arise during the inflationary era, and become a feature of
the primordial density fluctuations. The amount of non-gaussianity produced in
the simplest inflationary models is typically around a few per cent
\cite{Falk:1992sf,Gangui:1993tt,Acquaviva:2002ud},
and can be related to a standard set of slow roll parameters \cite{liddle}. For
comparison, the second order Sachs-Wolfe effect is expected to act as a source
of non-gaussianity in the cosmic microwave background observations equivalent
to $f_{NL}\sim 1$ \cite{Pyne:1995bs}.

Maldacena \cite{Maldacena:2002vr} introduced a simple argument which can be
used to determine the non-linearity parameter for cold inflation in the
`squeezed triangle limit' $k_1\ll k_2\approx k_3$. Density perturbations freeze
out, i.e. their amplitude becomes constant, when their wavelength exceeds the
Hubble length. Perturbations with the smallest wave
number $k_1$ freeze out first. Their effect on the bispectrum is equivalent
to  rescaling of the wave numbers $k_2$ and $k_3$ in the power spectrum.
The rescalling behaviour of the power spectrum is described by the spectral
index of scalar density perturbations $n$, leading to the result
\begin{equation}
f_{NL}={5\over 12}(n-1).\label{ma}
\end{equation}
This result is quite robust, and applies to many versions of inflation
\cite{Seery:2006js}. There are, however, reasons to be cautious when we
try to apply the same idea to warm inflationary models. In warm inflation, both
the radiation and the inflaton fluctuate \cite{Hall:2003zp}. The non-linear
coupling between these
fluctuations acts as a source of non-gaussianity, and this source is
impossible to describe in purely geometrical terms. Only one mode of
fluctuation survives after horizon crossing, but by then the non-gaussianity
is already imprinted on the curvature fluctuations.

The non-gaussianity produced by warm inflationary models has been looked at
previously by Gupta et al. \cite{Gupta:2002kn,Gupta:2005nh}. Our approach
builds upon such work, but now we include the nonlinear coupling between the
radiation and inflaton fluctuations on sub-horizon scales and find that a far
larger amount of non-gaussianity is produced. The contribution 
to the non-gaussianity found in the previous work was dependent on the third
derivative of the inflaton potential, equivalent to a second order effect
in the slow-roll parameter expansion. We shall see that there are large
contributions to the non-gaussianity, appearing at zeroth order in the
slow-roll approximation.

The best observational limit on the non-linearity function at present is from
the WMAP three-year data \cite{Spergel:2006hy}, which gives  $|f_{NL}|<114$.
The Planck satellite observations have a predicted sensitivity limit of around
$|f_{NL}|\sim5$ 
\cite{Komatsu:2001rj}. Our minimum prediction for $|f_{NL}|$ lies well above
the Planck threshold, with a distinctive angular dependence, and should provide
a means to test warm inflation observationally.

The paper is organised as follows. We begin in section II with a brief
introduction to the notion of warm inflation. In section III, we introduce
fluctuations of the inflaton field described by a Langevin equation. The
density fluctuations on large scales, which affect observations on the cosmic
microwave background are studied in section IV, followed by a discusion of some
of the consequences of our results, and extensions, in the conclusion. Appendix
A describes the first order perturbation theory relevant to warm inflation and
appendix B evaluates some of the integrals encountered in the main text.

\section{warm inflation}

Warm inflation occurs when there is a significant amount of particle production
during the inflationary era. We shall assume that
the particle interactions are strong enough to produce a thermal gas of
radiation with temperature $T$. Warm inflation is said to occur when $T$ is
larger than the energy scale set by the expansion rate $H$. The production of
radiation is associated with a damping effect on the inflaton, whose equation
of motion becomes
\begin{equation}
\ddot\phi+(3H+\Gamma)\dot\phi+V_\phi=0
\end{equation}
where $\Gamma(\phi,T)$ is a friction coefficient, $H$ is the Hubble parameter
and $V_\phi$ is the $\phi$ derivative of the inflaton potential $V(\phi,T)$.
Energy transfer to the radiation is associated with an increase of radiation
entropy density $s_r$ according to the equation
\begin{equation}
T\dot s_r+3HT s_r=\Gamma\dot\phi^2
\end{equation}
The effectiveness of warm inflation can be parameterised by a parameter $r$,
defined by
\begin{equation}
r={\Gamma\over 3H}
\end{equation}
When $r\gg 1$ the warm inflation is described as being in the strong regime.

Temperature dependence in the friction coefficient is a feature of many models
\cite{Moss:2006gt}, and can
lead to interesting effects on the density perturbations \cite{Hall:2003zp},
but in order to simplify the account given here we shall make the following
simplifications:
\begin{itemize}
\item $\Gamma\equiv \Gamma(\phi)$
\item $V\equiv V(\phi)$
\item $\Gamma\gg H$
\end{itemize}
In this case, the time evolution is described by the equations
\begin{eqnarray}
&&\ddot\phi+\Gamma\dot\phi+V_\phi=0,\\
&&\dot\rho_r+4H\rho_r=\Gamma\dot\phi^2,\\
&&3H^2=4\pi G\left(2V+2\rho_r+\dot\phi^2\right)
\end{eqnarray}
where $\rho_r$ is the radiation energy density.

During inflation we apply a slow-roll approximation and drop the highest
derivative terms in the equations of motion,
\begin{eqnarray}
&&\Gamma\dot\phi+V_\phi=0,\label{sr1}\\
&&4H\rho_r=\Gamma\dot\phi^2,\label{sr2}\\
&&3H^2=8\pi G V\label{sr3}
\end{eqnarray}
The validity of the slow-roll approximation depends on the slow roll parameters
defined in \cite{Hall:2003zp},
\begin{equation}
\epsilon={1\over 16\pi G}\left({V_\phi\over V}\right)^2,\qquad
\eta={1\over 8\pi G}\left({V_{\phi\phi}\over V}\right),\qquad
\beta={1\over 8\pi G}\left({\Gamma_\phi V_\phi\over \Gamma V}\right)
\end{equation}
The slow-roll approximation holds when $\epsilon\ll r$, $\eta\ll r$ and
$\beta\ll r$. Any quantity of order $\epsilon/r$ will be described as being
first order in the slow roll approximation.

\section{inflaton fluctuations}

Thermal fluctuations are the main source of density perturbations in warm
inflation. Thermal noise is transfered to the inflaton field mostly on small
scales. As the comoving wavelength of a perturbation expands, the thermal
effects decrease until the fluctuation amplitude freezes out. This may occur
when the wavelength of the fluctuation is still small in comparison with
cosmological scales.  

The behaviour of a scalar field interacting with radiation can be analysed
using the Schwinger-Keldysh approach to non-equilibrium field theory 
\cite{schwinger61,keldysh64}. When the small-scale behaviour of the 
interactions is averaged out, a simple picture emerges in which the field can
be described by a stochastic system evolving according to a Langevin
equation
\cite{calzetta88},
\begin{equation}
-\nabla^2\phi(x,t)+\Gamma\dot\phi(x,t)+V_\phi=\xi(x,t),\label{stoc}
\end{equation}
where $\nabla^2$ is the flat spacetime Laplacian and $\xi$ is a stochastic
source. For a weakly interacting gas with $T\gg\Gamma$, the source term has a
gaussian distribution with correlation function \cite{gleiser94}
\begin{equation}
\langle\xi(x,t)\xi(x',t')\rangle=2\Gamma T
\delta^{(3)}(x-x')\delta(t-t').
\end{equation}
We shall restrict ourselves to the gaussian noise source, although 
non-gaussianity in the noise could act as a source of 
non-gaussianity in the density fluctuations.

We can use the equivalence principle to adapt the flat spacetime Langevin
equation to an expanding universe with scale factor $a$. The rest frame of
the fluid will have a non-zero $3-$velocity $v_r^\alpha$ with respect to the
cosmological frame and we must include an advection term. The Langevin 
equation becomes
\begin{equation}
\ddot\phi(x,t)+\Gamma\dot\phi(x,t)+
\Gamma a^{-2}v_r^\alpha\partial_\alpha\phi(x,t)
+V_\phi-a^{-2}\partial^2\phi(x,t)=\xi(x,t)\label{langa}
\end{equation}
where $\partial^2$ is the Laplacian in an expanding frame with coordinates 
$x^\alpha$. The correlation function for the noise expressed in terms of the 
comoving cosmological coordinates becomes,
\begin{equation}
\langle\xi(x,t)\xi(x',t')\rangle=2\Gamma T a^{-3}(2\pi)^2
\delta^{(3)}(x-x')\delta(t-t').
\end{equation}
The inflaton will also generate metric inhomogeneities, but these can be small
on small scales. In first order perturbation theory, with a suitable choice of
gauge, the small scale metric fluctuations can be discarded on sub-horizon
scales (see appendix \ref{appa}). We will therefore apply eq. (\ref{langa}) on
sub-horizon scales and use a matching argument to extend the fluctuations to
large scales.

The stochastic equation for the inflaton field has some similarities to the one
which is used in the theory of stochastic inflation
\cite{Starobinsky:1986fx}, and we shall adopt a 
method for analysing the fluctuations which resembles one originally used by
Gangui et al. \cite{Gangui:1993tt} in that context. However, there are
important differences, reflecting the different
interpretation of the stochastic equation in the two applications. In
stochastic inflation, the large-scale inflaton field is constructed only from
modes which have wavelengths larger than the horizon, whereas the inflaton
field used here is valid for wavelengths larger than the scale of interactions
in the radiation, which is sub-horizon size. We therefore need to retain time
and spatial derivative terms which are dropped in Stochastic inflation.

We shall use a uniform expansion rate gauge. The analysis of the Langevin
equation can be simplified by introducing a new
time coordinate $\tau=(aH)^{-1}$ and using the slow roll approximation. We are
led to the equation
\begin{equation}
\phi^{\prime\prime}(x,\tau)-{3r\over \tau}\phi'(x,\tau)-
3r\,v_r^\alpha\partial_\alpha\phi(x,\tau)+{1\over (H\tau)^2}V_\phi
-\partial^2\phi(x,\tau)
=(2\Gamma T)^{1/2}\hat\xi(x,\tau)
\end{equation}
where a prime denotes a derivative with respect to $\tau$ and we have kept only
the leading terms in the slow roll approximation. The noise term has been
rescaled so that its correlation function is now
\begin{equation}
\langle\hat\xi(x,\tau)\hat\xi(x',\tau')\rangle=(2\pi)^3
\delta^{(3)}(x-x')\delta(\tau-\tau').\label{cf}
\end{equation}
This equation is non-linear because $\Gamma$, $v_r$ and $T$ depend on
$\phi$.

Now we treat the source term as a small perturbation and expand the inflaton
field
\begin{equation}
\phi(x,\tau)=\phi(\tau)+\delta_1\phi(x,\tau)+\delta_2\phi(x,\tau)+\dots
\label{ipe}
\end{equation}
where $\delta_1\phi$ is the linear response due to the source $\hat\xi$.
This expansion is substituted into the langevin equation and then we take the 
Fourrier transform. Only the zeroth order terms in the slow roll 
approximation will be retained. The equations for the first two inflaton 
perturbations are
\begin{eqnarray}
L\,\delta_1\phi
&=&(2\Gamma T)^{1/2}\hat\xi\label{del1}\\
L\,\delta_2\phi&=&
{1\over2T}(2\Gamma T)^{1/2}\delta_1T*\hat\xi
-3r\tau^{-1} (\hat k^\alpha v_r)*(k_\alpha \delta_1\phi)\label{del2}
\end{eqnarray}
where $v_r$ is the scalar velocity perturbation $v_r^\alpha=-ia\hat k^\alpha
v_r$, the operator $L$ is defined by
\begin{equation}
L\,\delta_1\phi=\delta_1\phi^{\prime\prime}-
{3r\over \tau}\delta_1\phi'+k^2\delta_1\phi
\end{equation}
and $*$ denotes the convolution
\begin{equation}
f*g({\bf k})=\int {d^3k'\over (2\pi)^3}f({\bf k}-{\bf k'})g({\bf k'}).
\end{equation}
Terms involving $\delta_1\Gamma$ have not been included because they appear at
first order in the slow roll expansion. Such terms produce non-gaussianity in
the fluctuations which depends on the first order slow roll parameters. A
useful consistency check can be performed to confirm that the terms which are
first order in the slow roll expansion reproduce Maldacena's result (\ref{ma})
in the squeezed triangle limit. Gupta et al \cite{Gupta:2002kn,Gupta:2005nh}
have analysed terms involving the third derivative of the potential
$V_{\phi\phi\phi}$. These
produce non-gaussianity which depends on second order slow
roll parameters.

The perturbation equations can be solved using green function techniques. The
solution to
$L\,u=j$ is
\begin{equation}
u={1\over k}\int_\tau^\infty G(k\tau,k\tau')(k\tau')^{1-2\nu}\,
j(k\tau')\,d\tau'.
\label{usol}
\end{equation}
The retarded green function $G(z,z')$ can be found for
$\nu=\Gamma/(2H)$ constant in terms of Bessel functions,
\begin{equation}
G(z,z')=-{\pi\over 2}z^{\nu} z^{\prime\nu}
\left(J_{\nu}(z)Y_{\nu}(z')-J_{\nu}(z')Y_{\nu}(z)\right).
\quad\hbox{ for }z<z'.
\label{rgf}
\end{equation}
Corrections due to the time dependence of $\nu$ are similar in size to terms
which we have already discarded in the slow roll approximation.

\subsection{Inflaton power spectrum}

The inflaton power spectrum $P_\phi(k,\tau)$ is defined by
\begin{equation}
\langle\delta\phi({\bf k}_1,\tau)\delta\phi({\bf k}_2,\tau)\rangle=
P_\phi(k_1,\tau)(2\pi)^3\delta({\bf k}_1+{\bf k}_2)
\end{equation}
Substituting the first order inflaton perturbation from (\ref{del1}), using the
general solution (\ref{usol}) and the correlation function (\ref{cf}) gives
\begin{equation}
P_\phi(k,\tau)=k^{-2}\int_\tau^\infty d\tau'G(k\tau,k\tau')^2(k\tau')^{2-4\nu}
\,2\Gamma(\tau')T(\tau')
\end{equation}
Integrals of this type are examined in appendix \ref{appb}. There is a saddle
point in the integral when $\nu$ is large, which allows us to take the $\Gamma
T$ out of the integral and obtain
\begin{equation}
P_\phi(k,\tau)=k^{-3}\,2\Gamma(\tau_F) \,T(\tau_F)\, F(k,\tau,\tau)
\end{equation}
where $\tau=\tau_F$ at the saddle point, or according to eq. (\ref{saddle}),
\begin{equation}
k=a\,\sqrt{3\over 2}(H\Gamma)^{1/2}.\label{freeze}
\end{equation}
We call the time in eq. (\ref{freeze}) the freezeout time $t_F$ for
the mode $k$. The freezeout time always precedes the horizon crossing time,
which occurs when
$k=aH$.

The remaining integral $F(k,\tau,\tau)$ is a special case of a more general
expression
\begin{equation}
F(k,\tau_1,\tau_2)=
k\int_{\tau_2}^\infty d\tau'
G(k\tau_1,k\tau')G(k\tau_2,k\tau')
(k\tau')^{2-4\nu}
\end{equation}
which is examined in appendix \ref{appb}. An analytic approximation of $F$
valid for large values of $\nu$ is given in
eq. (\ref{fapprox}). With this we recover a result derived in
\cite{Hall:2003zp},
\begin{equation}
P_\phi(k,\tau)=k^{-3}\,{\sqrt{\pi}\over2}(H\Gamma)^{1/2}T
\left(1+{H\over \Gamma}k^2\tau^2+\dots\right)\label{papprox}
\end{equation}
Note that $\tau$ decreases with time and $P_\phi(k,\tau)$ approaches a constant
value $P_\phi(k)$ on a timescale set by the freezout time. Horizon crossing
occurs at $\tau=1$.

\subsection{Inflaton bispectrum}

The bispectrum of the inflaton fields is defined by
\begin{equation}
\langle \delta\phi({\bf k}_1,\tau)
\delta\phi({\bf k}_2,\tau)
\delta\phi({\bf k}_3,\tau)\rangle=
(2\pi)^3B_\phi(k_1,k_2,k_3)
\delta^3({\bf k}_1+{\bf k}_2+{\bf k}_3).
\end{equation}
The first order inflaton perturbations $\delta_1\phi$ are gaussian fields and
their bispectrum vanishes. The leading order contribution to the bispectrum
must therefore include a contribution to $\delta\phi$ from the second order 
perturbation,
\begin{equation}
\sum_{\rm cyclic}\langle \delta_1\phi({\bf k}_1,\tau)
\delta_1\phi({\bf k}_2,\tau)
\delta_2\phi({\bf k}_3,\tau)\rangle\approx
(2\pi)^3B_\phi(k_1,k_2,k_3)
\delta^3({\bf k}_1+{\bf k}_2+{\bf k}_3),\label{bformula}
\end{equation}
where `cyclic' denotes cyclic permutations of 
$\{{\bf k}_1,{\bf k}_2,{\bf k}_3\}$. The second order perturbation can be
obtained by solving eq. (\ref{del2}). 
The most interesting effects arise from
the thermal fluctuations which are responsible for the temperature
perturbation $\delta_1T$ and velocity perturbations $v$. 

There are two types of fluctuation in the radiation. The first type is
the purely statistical type of fluctuation which is present due to the
microscopic particle motions. These fluctuations have been analysed before
\cite{Hall:2003zp}. They can be important, but they decrease rapidly after the
freezout time defined in the previous section. The second type of fluctuation
is driven by the energy flux from the inflaton field. These fluctuations are
important for coupling the inflaton and radiation fields, and we shall
consider these in more detail.

The fluctuations in the radiation field need only be evaluated to first order
in perturbation theory. The first order perturbation equations for the
complete system can be found in appendix \ref{appa}. The momentum flux $J$
from the decay of the inflaton field, given by
\begin{equation}
J=-\Gamma\dot\phi\,\delta_1\phi,\label{defj}
\end{equation}
is particularly important. If we keep only the leading terms in the slow roll
approximation,
the energy density and velocity perturbations in a uniform
expansion gauge satisfy the equations
\begin{eqnarray}
&&\delta_1\ddot\rho_{r}+9H\delta_1\dot\rho_{r}+
\left(20H^2+{1\over 3}k^2a^{-2}\right)\delta_1\rho_{r}=
k^2a^{-2}J\label{drho}\\
&&\delta_1\dot\rho_r+4H\delta_1\rho_r=-{4\over 3}ka^{-1}\rho_r\,v_r
\end{eqnarray} 
where the subscript $\kappa$ used in eq. (\ref{adrho}) to denote the gauge
choice
has been dropped.

We can solve eq. (\ref{drho}) exactly using green function methods as before,
\begin{eqnarray}
\delta_1\rho_r&=&\sqrt{3}\,k\,\tau^4\int_\tau^\infty\tau^{\prime-4}
\sin{k(\tau'-\tau)\over\sqrt{3}}J(\tau')\,d\tau'\\
v_r&=&-{3\over 4\rho_r}k\,\tau^4\int_\tau^\infty\tau^{\prime-4}
\cos{k(\tau'-\tau)\over\sqrt{3}}J(\tau')\,d\tau'
\end{eqnarray}
where $\tau=(aH)^{-1}$. We may obtain a reasonable approximation by taking
the slowly varying terms outside the integrals. Using eq. (\ref{defj})
and $\rho_r\propto T^4$, we have
\begin{eqnarray}
{\delta_1 T\over T}&=&-{3k\Gamma\dot\phi\over 4\rho_r}\,
\int_\tau^\infty g_r(k\tau,k\tau')\,
\delta_1\phi(\tau')d\tau'\label{drhor}\\
v_r&=&-{9\Gamma\dot\phi\over \tau \rho_r}\,
\int_\tau^\infty g_v(k\tau,k\tau')\,
\delta_1\phi(\tau')d\tau'\label{veq}
\end{eqnarray}
where,
\begin{eqnarray}
g_r(z,z')&=&{1\over \sqrt{3}}\left({z\over z'}\right)^4 
\sin{z'-z\over \sqrt{3}}\\
g_v(z,z')&=&{z\over 12}\left({z\over z'}\right)^4 
\cos{z'-z\over \sqrt{3}}
\end{eqnarray}
For later reference, we define the integral
\begin{equation}
g_5(z)=\int_z^\infty g_v(z,z')dz'=
{z^5\over 4!}\int_0^\infty{x^{4}e^{-zx}\over 1+3x^2}dx,
\label{gn}
\end{equation}

The fluctuations in the radiation dominate over all other terms which 
drive the second order inflaton fluctuations. We shall split the
second order inflaton perturbation into a part $\delta_v\phi$ driven by 
the fluid velocity and a part $\delta_r\phi$ driven by the thermal 
fluctuations,
\begin{equation}
\delta_2\phi=\delta_v\phi+\delta_r\phi
\end{equation}
The associated parts of the bispectrum will be denoted by $B_\phi^v$ and
$B_\phi^r$.
We can substitute the thermal fluctuations (\ref{drhor}) and (\ref{veq}) into 
eq. (\ref{del2}) and use the slow roll eqs. (\ref{sr1}-\ref{sr3}) to get
\begin{eqnarray}
L\,\delta_v\phi&=&A_v\,\tau^{-2}\int_\tau^\infty d\tau'
\int{d^3p\over (2\pi)^3}\,
g_v(p\tau,p\tau')\,
\hat {\bf p}\cdot({\bf k}-{\bf p})\,
\delta_1\phi({\bf p},\tau')\,\delta_1\phi({\bf k}-{\bf p},\tau)\\
L\,\delta_r\phi&=&A_r\,
\int_\tau^\infty d\tau'
\int{d^3p\over (2\pi)^3}\,
g_r(p\tau,p\tau')\,\delta_1\phi({\bf p},\tau')\,
\hat\xi({\bf k}-{\bf p},\tau)
\end{eqnarray}
where
\begin{equation}
A_r=-{3\over 2}{H\over\dot\phi},\qquad
A_v=36\,{\Gamma\over\dot\phi}
\end{equation}
The three-point function $B_\phi^v$ is given by substituting
the solution for $\delta_v\phi$ into eq. (\ref{bformula}),
\begin{eqnarray}
(2\pi)^3B_\phi^v(k_1,k_2,k_3)
\delta^3({\bf k}_1+{\bf k}_2+{\bf k}_3)&=&
\sum_{\rm cyclic}\int_\tau^\infty d\tau'\int_{\tau'}^\infty
d\tau^{\prime\prime}\int{d^3p\over (2\pi)^3}
\,\hat {\bf p}\cdot({\bf k}_3-{\bf p})\times\nonumber\\
&&G(k_3\tau,k_3\tau')(k_3\tau')^{1-2\nu}A_v(\tau')
\tau^{\prime-2}g_v(p\tau',p\tau^{\prime\prime})k_3^{-1}
\times\nonumber\\
&&\langle\delta_1\phi({\bf k}_1,\tau)\delta_1\phi({\bf k}_2,\tau)
\delta_1\phi({\bf p},\tau^{\prime\prime})
\delta_1\phi({\bf k}_3-{\bf p},\tau')\rangle.
\end{eqnarray}
The four-point function splits into a product of two-point functions. Using 
eq. (\ref{gn}) and the results from appendix \ref{appb}, we obtain an
approximation valid for large $\nu$,
\begin{equation}
B_\phi^v(k_1,k_2,k_3)\approx
\sum_{\rm cyclic}A_v\,P_\phi(k_1)P_\phi(k_2){{\bf k}_1\cdot {\bf k}_2\over k_3}
\int_\tau^\infty d\tau'
\left({g_5(k_1\tau')\over (k_1\tau')^2}
 +{g_5(k_2\tau')\over (k_2\tau')^2}\right)
G(k_3\tau,k_3\tau')(k_3\tau')^{1-2\nu}.
\end{equation}
From eq. (\ref{lastint}),
\begin{equation}
B_\phi^v(k_1,k_2,k_3)\approx18\,{H\over\dot\phi}
L(r)\sum_{\rm cyclic}
\left(P_\phi(k_1)P_\phi(k_2)(k_1^{-2}+k_2^{-2}){\bf k}_1\cdot {\bf k}_2
\right),
\label{bv}
\end{equation}
where $r=2\nu/3=\Gamma/(3H)$ and an asymptotic series for the function $L(r)$
is given
by eq. (\ref{Ldef}). 

A similar analysis for $B_\phi^r$ using eq. (\ref{fapprox}) gives,
\begin{equation}
B_\phi^r(k_1,k_2,k_3)\approx
3\,{H\over\dot\phi}\sum_{\rm cyclic}\left(
P_\phi(k_1)P_\phi(k_2)\left({2k_2^2\over k_2^2+k_3^2}\right)^{3/2}
\right).
\label{BT}
\end{equation}
This approximation.is valid for large $\nu$.

\section{Density fluctuations}

The small scale inflaton fluctuations freeze out well in advance of the time
when they cross the horizon. Whilst these fluctuations are freezing out, the
metric fluctuations are relatively small (in the uniform expansion rate gauge:
see appendix \ref{appa}). This has two important consequences. In the first
place, we are relieved of the arduous task of doing second order
perturbation theory for the metric perturbations. Secondly, we do not have to
consider the effects of metric perturbations in the thermal field theory used
to obtain the
Langevin equation (\ref{langa}). By contrast, in Maldacena's analysis of
non-gaussianity in cold inflation \cite{Maldacena:2002vr}, it was necessary to
consider the
inflaton vacuum fluctuations in conjunction with metric perturbations because
the second order perturbations where the same order in the slow roll parameters
as the metric perturbations and arose on horizon scales.

Eventually, the wavelength of the perturbations crosses the effective
cosmological horizon the metric perturbations become important. On large scales
it becomes possible to use a small-spatial-gradient expansion (first
formalised by Salopeck and Bond \cite{Salopek:1990jq}). This approach allows
us to define the curvature perturbation $\zeta$ so that it is conserved even
in the non-linear theory 
\cite{Sasaki:1995aw,Lyth:2004gb,Lyth:2005fi}. 
The bispectrum  and the 
non-linearity of the density fluctuations can be obtained, to a reasonable
accuracy, by matching the small and large scale approximations at horizon
crossing. 

The large scale behaviour is governed by the same equations as the homogeneous
system. In particular, during inflation the slow roll equations can be used to
relate the total pressure and density to the value of the inflaton field. In
this situation, the fluctuations can be described entirely by the conserved
expansion fluctuation $\zeta$ on constant density hypersurfaces, which is
defined for general hypersurfaces by
\begin{equation}
\zeta=\frac12\ln\left(1+2\varphi\right)+\int{d\rho\over p+\rho},
\end{equation}
where $\varphi$ is the spatial curvature perturbation. After 
using the slow roll equations,
\begin{equation}
\zeta=\frac12\ln\left(1+2\varphi\right)+\int {H\over\dot\phi}d\phi,
\end{equation}
where $H\equiv H(\phi)$ and $\dot\phi\equiv \dot\phi(\phi)$ are given from eqs.
(\ref{sr1}-\ref{sr3}).

Consider a uniform curvature gauge $\varphi=0$ and
$\zeta\equiv\zeta(\phi)$. When the inflaton perturbations are expanded as
before in eq. (\ref{ipe}), we have 
\begin{equation}
\zeta=\zeta_\phi\delta_1\phi+
+\zeta_\phi\delta_2\phi+
\frac12\zeta_{\phi\phi}\delta_1\phi*\delta_1\phi+\dots
\end{equation}
where $\phi$ subscripts denote derivatives with respect to $\phi$ and 
$\zeta_\phi=H/\dot\phi$. Hence the power spectrum of density
perturbations is
\begin{equation}
P_\zeta(k)=\zeta_\phi^2 P_\phi(k)
\end{equation}
where $P_\phi(k)$ is evaluated in a uniform curvature gauge. The bispectrum,
using eq. (\ref{bformula}), is given by
\begin{equation}
B_\zeta(k_1,k_2,k_3)=
\zeta_\phi^3 B_\phi(k_1,k_2,k_3)+
{\zeta_{\phi\phi}\over\zeta_\phi^2}(P_\phi(k_1)P_\phi(k_2)+\hbox{cyclic}).
\label{bzeta}
\end{equation}
Note that, according to eqs. (\ref{sr1}-\ref{sr3}), 
$\zeta_{\phi\phi}/\zeta^2_\phi$ is first order in the slow roll expansion.
This term is important for relating the non-gaussianity to the slow roll 
parameters, but in our case the non-gaussianity occurs at zeroth order 
in the slow roll expansion and we can neglect terms which are first order. 

In the previous section we evaluated the small scale inflaton perturbations
in a uniform expansion-rate gauge. On sub-horizon scales, we have argued
already that the metric perturbations are relatively small and we should
expect that the inflaton perturbations in the uniform expansion-rate gauge and
the uniform curvature gauge should be approximately equal. This can be checked
explicitly at first order in perturbation theory using eq. (\ref{phivarphi})
in appendix \ref{appa}. We will therefore use our earlier results for the
inflaton fluctuations when applying eq. (\ref{bzeta}).

\begin{center}
\begin{figure}[ht]
\scalebox{0.5}{\includegraphics{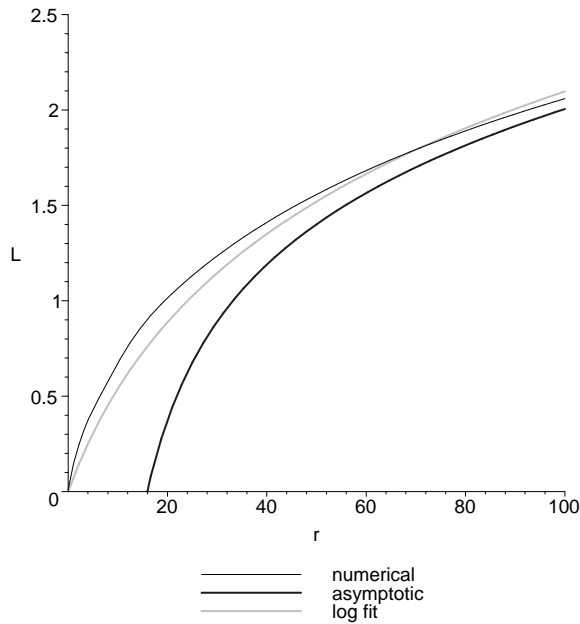}}
\caption{The function $L(r)$ determines the size of the non-gaussianity. The
upper curve is a numerical calculation and the lower an asymptotic
approximation (\ref{Ldef}) valid for large $r$. A log-linear fit 
eq. (\ref{logfit}) is also shown. }
\label{figfnl}
\end{figure}
\end{center}

The density fluctuation bispectrum can now be obtained by matching the large
scale curvature bispectrum eq. (\ref{bzeta}) to the small scale inflaton
bispectrum eqs. (\ref{bv}) and (\ref{BT}). The largest contribution to the
bispectrum, which we denote by $B_\zeta^v$, comes from the velocity term
$B_\phi^v$,
\begin{equation}
B_\zeta^v(k_1,k_2,k_3)\approx18\,
L(r)\sum_{\rm cyclic}
\left(P_\zeta(k_1)P_\zeta(k_2)(k_1^{-2}+k_2^{-2}){\bf k}_1\cdot {\bf k}_2
\right),
\end{equation}
which is valid for large $r=\Gamma/(3H)$, where $L(r)$ is plotted in Figure
\ref{figfnl}.

\begin{center}
\begin{figure}[ht]
\scalebox{0.5}{\includegraphics{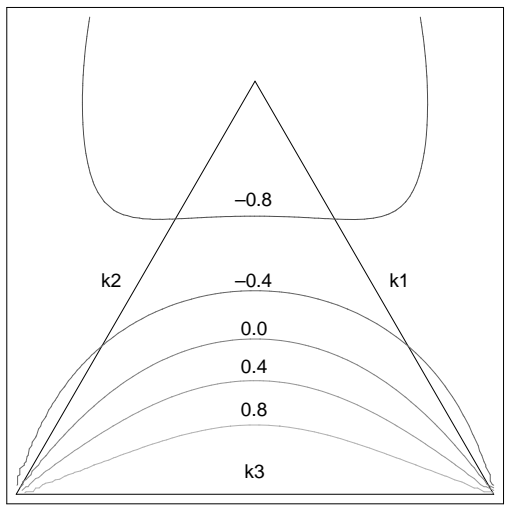}}
\caption{This contour plot shows the momentum dependence of $f_{NL}^v$. The
momentum $k_3$ is held fixed and placed along the $x-$ axis, whilst $k_2$ and
$k_3$ are allowed to vary over the boxed region. The contour values must be
multiplied by $15L(r)$ to obtain $f_{NL}^v$.}
\label{figfk}
\end{figure}
\end{center}

The amount of non-gaussianity contained in this part of the bispectrum is
described by the non-linearity function $f^v_{NL}$, defined as in eq.
(\ref{fdef}). If we take a scale free spectrum with $P_\zeta(k)\propto
k^{-3}$, then  
\begin{equation}
f_{NL}^v(k_1,k_2,k_3)=
{15\,L(r)\over k_1^3+k_2^3+k_3^3}
\left({k_3^3(k_1^2+k_2^2){\bf k}_1\cdot {\bf k}_2\over k_1^2k_2^2}
+{k_1^3(k_2^2+k_3^2){\bf k}_2\cdot {\bf k}_3\over k_2^2k_3^2}+
{k_2^3(k_3^2+k_1^2){\bf k}_3\cdot {\bf k}_1\over k_3^2k_1^2}\right),
\label{fv}
\end{equation}
The value of $f_{NL}^v$ always lies in
the range $-15 L(r)<f_{NL}^v<(33/2)L(r)$. In the equilateral triangle limit,
\begin{equation}
f_{NL}^v(k,k,k)=-15\,L(r).
\end{equation}
The momentum dependence of the bispectrum contributes to the angular bispectrum
in the cosmic microwave backround. This would be an important signature of
warm inflation. Figure \ref{figfk} showns the momentum dependence of the
non-linearity function $f_{NL}^v$  on a contour plot.

The derivation of our result may fail in the squeezed triangle limit,
when $k_1\ll k_2\approx k_3$,  because one of the modes may exit the horizon
before the other modes have frozen out. The freezeout time is
given by eq. (\ref{freeze}), so that the consistency condition on our result
is that
\begin{equation}
{k_1\over k_2}>\left({H\over \Gamma}\right)^{1/2}.
\end{equation}
We should also be prepared for the possibility that the time evolution of $r$
during the inflationary era gives an additional small dependence on $k$.

The contribution from the radiation density
fluctuation terms, which we denote by $f_{NL}^r$, is given by
\begin{equation}
f_{NL}^r(k_1,k_2,k_3)=
-{5\over 2}{1\over k_1^3+k_2^3+k_3^3}
\left\{\left({2k_1^2k_2^2\over k_1^2+k_2^2}\right)^{3/2}+
\left({2k_2^2k_3^2\over k_2^2+k_3^2}\right)^{3/2}+
\left({2k_3^2k_1^2\over k_3^2+k_1^2}\right)^{3/2}\right\}
\end{equation}
This is substantially smaller than $f_{NL}^v$, but has the advantage that 
it is independent of the parameters in the model. In the squeezed triangle
limit, $f_{NL}^r=-5/4$, which is identical to the `generic' result for the
non-gaussianity produced by the curvaton model \cite{Lyth:2005fi}.

\section{conclusion}

We have presented a preliminary analysis of the amount of non-gaussianity in
the density fluctuations in the warm inflationary scenario. The discussion has
been restricted to the strong, thermal regime of warm inflation where the
radiation produced from the inflaton vacuum energy thermalises and where there
is a large friction coefficient $\Gamma\equiv\Gamma(\phi)$ in the inflaton
equation of motion. We have found that the interaction between inflaton and
radiation fluctuations leads to a large non-gaussianity in the density
fluctuations when compared to cold single field inflation.

The amount of non-gaussianity in the bispectrum, measured by the non-linearity
function $f_{NL}$ with equal momenta, is given approximately by
\begin{equation}
f_{NL}\approx -15\ln\left(1+{r\over 14}\right)-{5\over 2},
\end{equation}
where $r=\Gamma/(3H)$ is the parameter which must be large 
for the strong type of warm inflation. There are no simple means of 
measuring  the value of $r$ from observations apart from this effect on the 
non-gaussianity. A naive use of the limit on the non-gaussianity parameter from
the WMAP three-year data  \cite{Spergel:2006hy}, $|f_{NL}|\le 114$, 
would correspond to $r\le2.8\times 10^4$. 

The sensitivity of observations by the Planck satellite has been estimated to
be around $|f_{NL}|\sim 5$, although this assumes that the primordial value of 
$|f_{NL}|$ is independent of momentum \cite{Komatsu:2001rj}. A limit
$|f_{NL}|< 5$, which might arise if Planck does not detect any primordial
non-gaussianity, would correspond to $r\le2.5$ and would not be
compatible with the strong  version of warm inflation which we have assumed
here. 

If warm inflation is realised in the way which we have supposed, then the
prospects for observing primordial non-gaussianity in the cosmic microwave
background are very good. The angular bispectrum of the cosmic microwave
background could then be used to distinguish warm inflation from other causes
of non-gaussianity, for example special cases of the curvaton scenario
\cite{Lyth:2005fi,Sasaki:2006kq},  which can also produce a significant amount
of non-gaussianity. A useful way to parameterise the bispectrum would be to
take
\begin{equation}
B_\zeta(k_k,k_2,k_3)=
\sum_{\rm cylic}\sum_{l=0}^{\rm lmax}P_\zeta(k_1)P_\zeta(k_2)
f_l(k_1,k_1)P_l(\hat{\bf k}_1\cdot\hat{\bf k}_2),
\end{equation}
where $P_l$ is a Legendre polynomial. Our result for warm inflation corresponds
to the $l=1$ term with $f_1$ linear in $k$. The curvaton scenario produces a
non-linearity $f_0=f_{NL}$ constant. If the angular dependence of the signal
corresponded to the dominance of the $l=1$ term, this would provide very
strong  evidence indeed in support of the warm inflation scenario. 

Further work can be done to generalise the results obtained here to
more general types of warm inflation. One
possible extension would be to consider temperature dependence in the
friction term, i.e. $\Gamma(\phi,T)$. Numerical work in \cite{Hall:2003zp} 
shows that this increases the amount of interaction between the inflaton and 
temperature fluctuations, and therefore we might expect even more 
non-gaussianity to develop. So far, the behaviour of the perturbations
in this situation is only understood numerically, and further analytic
work would be desirable for predictions of the non-gaussianity. 

For a complete picture of the non-gaussianity in models of warm inflation, 
we should also consider the weak regime of warm inflation, where
the friction coefficient is small. In the weak regime, the fluctuations
freeze out at horizon crossing. This invalidates most of our analytic results,
but it should still be possible to evaluate the necessary integrals
numerically.

\appendix

\section{First order cosmological perturbations}\label{appa}

In this appendix we shall review the equations for the first order cosmological
perturbations of an inflaton and radiation system. These where first
constructed in refs. \cite{Lee:2000dt,DeOliveira:2001he}. Our
notation closely follows the review by Hwang and Noh \cite{hwang02}.

The most general scalar perturbations of the metric can be written in the form
\begin{equation}
ds^2=-(1+2\alpha)dt^2-2a^2\beta_{,\alpha}dtdx^\alpha+
a^2(\delta_{\alpha\beta}(1+2\varphi)+2\gamma_{,\alpha\beta})dx^\alpha dx^\beta
\end{equation}
where $a(t)$ is the scale factor and $\alpha$, $\beta$, $\varphi$ and $\gamma$
depend on space and time. We always use a Fourier transform with wave vector
$k_\alpha$ to replace the dependence on $x^\alpha$. We will also find it
useful to introduce the perturbed expansion $\kappa$ and the shear $\chi$ of
the normals to the time slices. Their spatial Fourier transforms are related
to the metric
fluctuations by
\begin{eqnarray}
\kappa&=&3(-\dot\varphi+H\alpha)+k^2a^{-2}\chi\\
\chi&=&a(\beta+a\dot\gamma)
\end{eqnarray}

The energy momentum tensor splits into a radiation part $T_{r\,ab}$,
and an inflaton part $T_{\phi\,ab}$, given by
\begin{eqnarray}
T_{r\,ab}&=&(\rho_r+p_r)v_{ra}v_{rb}+p_rg_{ab}\\
T_{\phi\,ab}&=&\phi_{,a}\phi_{,b}-(\frac12g^{cd}\phi_{,c}\phi_{,d}+V)g_{ab}
\end{eqnarray}
where the $\rho_r$, $p_r$ and $v_{ra}$ are the radiation density, pressure and
$4-$velocity. Energy transfer between the two components is described by a
flux term $Q_a$,
\begin{equation}T_{rab}{}^{;b}=Q_a\end{equation}
The inflaton equation of motion given earlier is equivalent to the choice
\begin{equation}
Q_a=-\Gamma v_r^b\phi_{,b}\phi_{,a}\label{q1}
\end{equation}

In the unperturbed system, the values of the total density $\rho$, pressure
$p$ and all other background quantities depend only on time. Perturbations to
the density and velocity are defined by
\begin{equation}\delta T_{\hat0\hat0}=\delta\rho,\qquad 
\delta T_{\hat0\hat\alpha}=i\hat k_{\hat\alpha} (p+\rho)v\end{equation}
where the orthonormal basis is denoted by hats, and $\hat k_\alpha$ is the
normalised wave vector. Perturbations to the radiation are defined
in a similar way in terms of $T_{r\,ab}$. The perturbed density and pressure
become
\begin{eqnarray}
\delta\rho
&=&\delta\rho_r+\dot\phi\delta\dot\phi-\dot\phi^2\alpha+V_\phi\delta\phi\label{r1}\\
\delta p&=&
w\delta\rho_r+\dot\phi\delta\dot\phi-\dot\phi^2\alpha-V_\phi\delta\phi\label{p1}
\end{eqnarray}
where $w=1/3$ and $\delta\phi$ is the inflaton perturbation.
Perturbations to the energy momentum transfer are described by the energy
transfer $\delta Q$ and momentum flux $J$,
\begin{equation}\delta Q_{\hat 0}=-\delta Q,\qquad \delta
Q_{\hat\alpha}=ik_{\hat\alpha} J\end{equation}
The explicit expressions obtained from eq. (\ref{q1}) are
\begin{eqnarray}
\delta Q&=&\delta\Gamma\dot\phi^2+
2\Gamma\dot\phi\delta\dot\phi-2\alpha\Gamma\dot\phi^2\\
J&=&-\Gamma\dot\phi\,\delta\phi
\end{eqnarray}

The relevant first order Einstein equations become
\begin{eqnarray}
&&k^2a^{-2}\varphi-H\kappa=4\pi G\,\delta\rho\label{a1}\\
&&\kappa-k^2 a^{-2}\chi=12\pi G\, k^{-1} a(\rho+p)v\label{a2}\\
&&\dot\chi+H\chi-\alpha-\varphi=0\label{a3}\\
&&\dot\kappa+2H\kappa+(3\dot H-k^2a^{-2})\alpha=
4\pi G(\delta\rho+3\delta p)\label{a4}
\end{eqnarray}
The inflaton equation becomes
\begin{equation}
\delta\ddot\phi+3H\delta\dot\phi+k^2 a^{-2}\delta\phi+V_{\phi\phi}=
\dot\phi(\kappa+\dot\alpha)+(2\ddot\phi+3H\dot\phi)\alpha-\delta Q
\label{a5}
\end{equation}
The energy and momentum equations for the radiation become
\begin{eqnarray}
\delta\dot\rho_r+3H(1+w)\delta\rho_r&=&-k a^{-1}(1+w)v_r+\dot\rho_r\alpha
+(1+w)\rho_r\kappa+\delta Q\label{a6}\\
(a^4\rho_r v_r)^\cdot&=& ka^3\rho_r \alpha+k a^3(1+w)^{-1}(w\delta\rho_r-J)
\label{a7}
\end{eqnarray}
Only six of the seven equations (\ref{a1}-\ref{a7}) are independent. There are
seven functions $\varphi$, $\kappa$, $\chi$, $\alpha$, $v_r$, $\delta\rho_r$
and $\delta\phi$, but the  gauge freedom $t\to t+\zeta(x)$ allows us to choose
the functional form any one of them at will.

The choice of gauge is rather arbitrary, but for short-wavelength calculations
where $ak>H$, the uniform expansion-rate gauge $\kappa=0$ proves to be
convenient.
This gauge choice is indicated by a subscript $\kappa$. If we eliminate the
fluid velocity $v_{r\kappa}$,
\begin{eqnarray}
&&\delta\ddot\rho_{r\kappa}+(8+3w)H\delta\dot\rho_{r\kappa}+
(3(1+w)\dot H+15(1+w)H^2+wk^2a^{-2})\delta\rho_{r\kappa}=\nonumber\\
&&\qquad k^2a^{-2}(J_\kappa-(1+w)\rho_r\alpha_\kappa)+
a^{-5}\left(a^5(\delta Q_\kappa+\dot\rho_r\alpha_\kappa)\right)^\cdot
\label{adrho}
\end{eqnarray}
where $\alpha_\kappa$ is given by substituting eqs. (\ref{r1}) and (\ref{p1})
into eq. (\ref{a4}),
\begin{equation}
(3\dot H-k^2a^{-2}+16\pi G\dot\phi^2)\alpha_\kappa=
4\pi G((1+3w)\delta\rho_{r\kappa}+4\dot\phi\delta\dot\phi_\kappa-
2 V_\phi\delta\phi_\kappa).
\end{equation}
In the uniform expansion-rate gauge, the metric perturbation 
$\alpha_\kappa$ drops out of eqs. (\ref{a5}) and 
(\ref{adrho}) when we work to leading order in the slow roll approximation. 
This was implicitly assumed in the discussion of the stochastic inflaton 
equation in section III. 

Converting the perturbations to other gauges can be done by examining gauge 
invariant combinations. For example, the combination
\begin{equation}
\delta\phi_{\varphi}=
\delta\phi-{\dot\phi\over H}\varphi
\end{equation}
is gauge invariant and represents the inflaton fluctuation in a constant
curvature gauge. Hence, in terms of uniform expansion-rate quantities, 
\begin{equation}
\delta\phi_\varphi=\delta\phi_\kappa
-{\dot\phi\over H}\varphi_\kappa.
\end{equation}
We can substitute for $\phi_\kappa$ from eq. (\ref{a1}), and obtain the
equation
\begin{equation}
\delta\phi_\varphi=\delta\phi_\kappa
-{4\pi Ga^2\dot\phi\over Hk^2}\delta\rho_\kappa\label{phivarphi}.
\end{equation}
Note that, during inflation, $\delta\rho_\kappa\approx V_\phi\delta\phi_\kappa$
and we find that $\delta\phi_\varphi\approx\delta\phi_\kappa$ to leading order
in the slow roll approximation.

\section{Integrals}\label{appb}

We begin with an approximation to the integral
\begin{equation}
F(k,\tau_1,\tau_2)=
k\int_{\tau_2}^\infty d\tau'
G(k\tau_1,k\tau')G(k\tau_2,k\tau')
(k\tau')^{2-4\nu}
\end{equation}
where $\nu=\Gamma/2H$ and the retarded green function $G$ is given in eq.
(\ref{rgf}). The leading terms for large $\nu$ and fixed $\tau_1$, $\tau_2$
come from
\begin{equation}
F(k,\tau_1,\tau_2)\approx
{\pi^2\over 4}k^3
\tau_1^\nu\tau_2^\nu Y_{\nu}(k\tau_1)Y_{\nu}(k\tau_2)
\int_0^\infty J_{\nu}(k\tau')^2\tau^{\prime 2-2\nu}d\tau'
\end{equation}
This is a standard integral,
\begin{equation}
\int_0^\infty J_{\nu}(k\tau')^2
\tau^{\prime 2-2\nu}d\tau'=
{k^{2\nu}2^{-1-2\nu}\sqrt{2\pi}
\over\Gamma_R(\nu+1)\Gamma_R(\nu-1/2)}
\end{equation}
where $\Gamma_R$ is the gamma function. We also have
\begin{equation}
z^\nu Y_\nu(z)\sim 
-{2^\nu\over\pi\Gamma_R(\nu)}\left(1+{z^2\over 4\nu}+\dots\right).
\end{equation}
Hence, 
\begin{equation}
F(k,\tau_1,\tau_2)\sim
\sqrt{\pi\over 32\nu}
\left(1+{k^2(\tau_1^2+\tau_2^2)\over 4\nu}+\dots\right).\label{fapprox}
\end{equation}

A slightly more chalenging problem is the integral
\begin{equation}
F(k_1,k_2,\tau,g)=
(k_1k_2)^{1/2}\int_\tau^\infty 
G(k_1\tau,k_1\tau')G(k_2\tau,k_2\tau')
(k_1\tau')^{1-2\nu}(k_2\tau')^{1-2\nu}g(\tau')\,d\tau'
\end{equation}
where $g(\tau)$ is a smooth function. We proceed as above to get 
\begin{equation}
F(k_1,k_2,\tau)\approx
{\pi^2\over 4}(k_1k_2)^{3/2}\tau^{2\nu}Y_{\nu}(k_1\tau)Y_{\nu}(k_2\tau)
\int_0^\infty J_{\nu}(k_1\tau')J_{\nu}(k_2\tau')
\tau^{\prime 2-2\nu}g(\tau')d\tau'.
\end{equation}
For large $\nu$ there is a Debye approximation for the Bessel functions which
is valid in the range $0<\tau<\nu(k_1k_2)^{-1/2}$, 
\begin{equation}
J_\nu(k_i\tau)\sim (2\pi\nu\tanh\alpha_i)^{-1/2}e^{\nu(\tanh\alpha_i-\alpha_i)}
\end{equation}
where $\nu\,{\rm sech}\alpha_i=k_i\tau'$ for $i=1,2$. The relevant part of the
integral becomes
\begin{equation}
F(k_1,k_2,\tau)\approx
{\pi\over 8}(k_1k_2)^{\nu}\tau^{2\nu}Y_{\nu}(k_1\tau)Y_{\nu}(k_2\tau)
\nu^{2-2\nu}\int_0^\infty(\cosh\alpha)^{2\nu-3}
\,e^{\nu(\tanh\alpha_1-\alpha_1+
\tanh\alpha_2-\alpha_2)}\,g(\tau')\,d\alpha
\end{equation}
where $\nu{\rm sech}\alpha=(k_1k_2)^{1/2}\tau'$. We find that there is a saddle
point in this range at the value of $\alpha$ corresponding to $\tau=\tau_F$,
where
\begin{equation}
\tau_F=(6\nu)^{1/2}(k_1^2+k_2^2)^{-1/2}\label{saddle}
\end{equation}
Expanding about the saddle point gives
\begin{equation}
F(k_1,k_2,\tau,g)\sim
\sqrt{\pi\over 4\nu}\left({k_1k_2\over k_1^2+k_2^2}\right)^{3/2}g(\tau_F)
\end{equation}
which agrees with our earlier result when $g(\tau)=1$. This saddle point
is responsible for the phenomenon of `freezing out' of the thermal
fluctuations. The value of $\tau$ decreases with time and the fluctuations
always freeze out before they cross the horizon at $\tau=1$.

One final integral which we require is
\begin{equation}
F_v(k_1,k_2,\tau)=
k_1\int_\tau^\infty d\tau' G(k_1\tau,k_1\tau')(k_1\tau')^{1-2\nu}
g(k_2\tau'),
\end{equation}
where
\begin{equation}
g(z)={1\over 4!}\int_0^\infty {x^4 e^{-x}\over z^2+3x^2}dx.
\end{equation}
The leading order behaviour for large $\nu$ is given by
\begin{equation}
F_v(k_1,k_2,\tau)\sim
(k_1\tau)^\nu Y_\nu(k_1\tau){\pi\over 2}
\int_0^\infty k_1d\tau'\,J_\nu(k_1\tau') (k_1\tau')^{1-\nu}g(k_2 \tau')
\end{equation}
We could use a saddle point approximation, and this produces a good
approximation when $g(z)$ is replaced by a constant. However, the integrand is
too flat to produce a good enough approximation with the function $g(z)$.
Instead, we interchange the orders of integration and use the identity
\begin{eqnarray}
&&\int_0^\infty dz\, J_\nu(z)z^{1-\nu}(z^2+3x^2)^{-1-\mu}=\nonumber\\
&&{(3x^2)^{-\mu}\Gamma_R(\mu)\over 
2^{1+\nu}\Gamma_R(1+\nu)\Gamma_R(1+\nu)}
\,{}_1F_2(1;1-\mu,1+\nu;3x^2/4)\nonumber\\
&&+{\Gamma_R(-\mu)\over 
2^{1+\nu+2\mu}\Gamma_R(1+\nu+\mu)}
\,{}_1F_2(1+\mu;1+\mu,1+\nu+\mu;3x^2/4)
\end{eqnarray}
We can integrate over $x$, let $\mu\to 0$ and use the large $\nu$ limit, to get
the asymptotic expansion
\begin{equation}
F_v(k_1,k_2,\tau)\sim \left(k_1\over k_2\right)^2
{1\over 4\nu}L\left({2\nu\over 3}\right)-
\left(k_1\over k_2\right)^2
\ln\left({k_1\over k_2}\right){1\over 2\nu}.\label{lastint}
\end{equation}
where
\begin{equation}
L(r)\sim \ln{2r}-{25\over 6}+\gamma+{30\over r}\ln{2r}
-{121\over r}+\dots\label{Ldef}
\end{equation}
and $\gamma$ is Euler's constant $0.57721\dots$. The asymptotic expansion
is not very accurate for moderate values of $r$, and a numerical evaluation
is provided in fig \ref{figfnl}. We can use a logarithmic fit,
\begin{equation}
L(r)\approx \ln\left( 1+{r\over 14}\right)\label{logfit}
\end{equation}
for rough estimates.

\bibliography{paper.bib,cosper.bib}

\end{document}